\documentclass[letterpaper]{article} % DO NOT CHANGE THIS
\usepackage{aaai2026}  % DO NOT CHANGE THIS
\usepackage{times}  % DO NOT CHANGE THIS
\usepackage{helvet}  % DO NOT CHANGE THIS
\usepackage{courier}  % DO NOT CHANGE THIS
\usepackage[hyphens]{url}  % DO NOT CHANGE THIS
\usepackage{graphicx} % DO NOT CHANGE THIS
\usepackage{natbib}  % DO NOT CHANGE THIS AND DO NOT ADD ANY OPTIONS TO IT
\usepackage{caption} % DO NOT CHANGE THIS AND DO NOT ADD ANY OPTIONS TO IT
\usepackage{algorithm}
\usepackage{algorithmic}
\usepackage{booktabs}   
\usepackage{multirow}  
\usepackage{amsmath}   
\usepackage{amssymb}    
\usepackage{cite}
\usepackage[flushleft]{threeparttable}
\usepackage{newfloat}
\usepackage{listings}
\DeclareCaptionStyle{ruled}{labelfont=normalfont,labelsep=colon,strut=off} % DO NOT CHANGE THIS
\floatstyle{ruled}
\newfloat{listing}{tb}{lst}{}
\floatname{listing}{Listing}
\title{Melodia: Training-Free Music Editing Guided by Attention\\ Probing in Diffusion Models}
\author{
    Yi Yang\footnote{These authors contributed equally.},
    Haowen Li\footnotemark[1],
    Tianxiang Li,
    Boyu Cao,
    Xiaohan Zhang,
    Liqun Chen,
    Qi Liu\thanks{Corresponding author.}
}
\affiliations{
    School of Future Technology, South China University of Technology\\
    ftyy@mail.scut.edu.cn, fthaowen@mail.scut.edu.cn, drliuqi@scut.edu.cn
}

\usepackage{bibentry}
\begin{document}

\maketitle

\begin{abstract}
Text-to-music generation technology is progressing rapidly, creating new opportunities for musical composition and editing. However, existing music editing methods often fail to preserve the source music's temporal structure, including melody and rhythm, when altering particular attributes like instrument, genre, and mood. To address this challenge, this paper conducts an in-depth probing analysis on attention maps within AudioLDM 2, a diffusion-based model commonly used as the backbone for existing music editing methods. We reveal a key finding: cross-attention maps encompass details regarding distinct musical characteristics, and interventions on these maps frequently result in ineffective modifications. In contrast, self-attention maps are essential for preserving the temporal structure of the source music during its conversion into the target music. Building upon this understanding, we present Melodia, a training-free technique that selectively manipulates self-attention maps in particular layers during the denoising process and leverages an attention repository to store source music information, achieving accurate modification of musical characteristics while preserving the original structure without requiring textual descriptions of the source music. Additionally, we propose two novel metrics to better evaluate music editing methods. Both objective and subjective experiments demonstrate that our approach achieves superior results in terms of textual adherence and structural integrity across various datasets. This research enhances comprehension of internal mechanisms within music generation models and provides improved control for music creation.
\end{abstract}


%     \link{Extended version}{https://aaai.org/extended-version}

\begin{figure}
    \label{fig1-challenge}
    \centering
    \includegraphics[width=\linewidth]{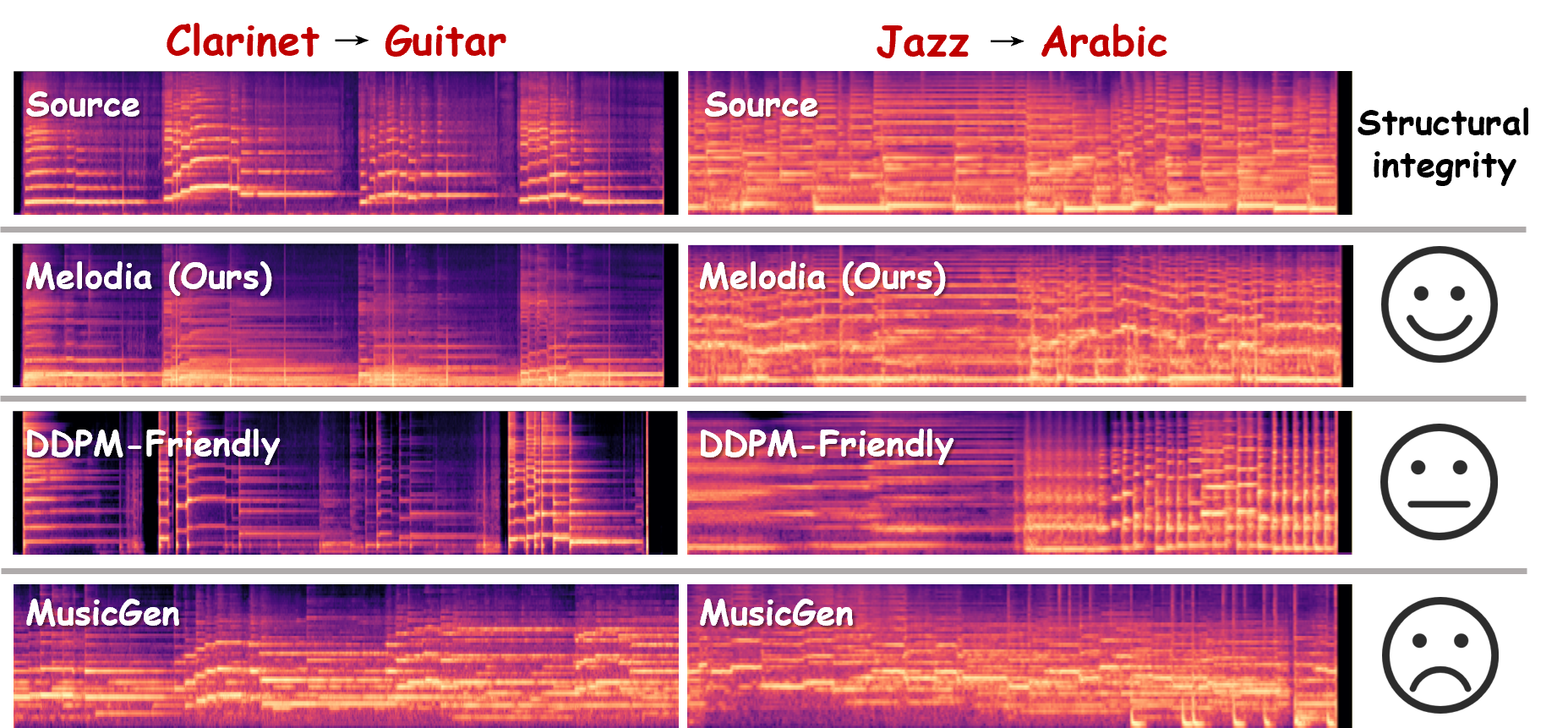}
    \caption{Spectrogram comparison of music editing results between different methods. The comparison reveals that existing methods struggle to preserve the original temporal patterns and rhythmic structures, while Melodia maintains better structural consistency with the source music.}
    \label{fig:first}
\end{figure}

\section{INTRODUCTION}

Text-to-music generation technology continues to evolve rapidly, opening new ways to edit music. Text-guided music editing enables modification of musical attributes through natural language instructions. As mentioned by MusicMagus~\cite{magus}, this field encompasses two categories: inter-stem editing and intra-stem editing. Inter-stem editing involves adding or removing separate instrumental tracks (e.g., "add drum" or "remove guitar"), while intra-stem editing focuses on modifying characteristics within a single track, such as changing the timbre, style, or mood while maintaining its melody and structure.

However, existing text-guided music editing methods struggle with these issues: 1) Existing methods train specialized models from scratch~\cite{agostinelli2023musiclm,musicgen} or fine-tune pre-trained models~\cite{ruiz2023dreambooth,zhang2024instructmusicgen}, both requiring significant computational costs and training data. 2) Most existing methods require textual descriptions of the source music to guide the editing process. For example, MusicMagus~\cite{magus} requires a descriptive word about the source music, which can be difficult for users to provide accurately. 3) Existing methods often struggle to maintain the temporal structure, including the melody and rhythm of the source music during editing. Fig.~\ref{fig:first} demonstrates this issue through spectrogram comparisons, where methods like DDPM-Friendly~\cite{manor2024ddpm} and MusicGen~\cite{musicgen} show inadequate structural preservation. 4) While attention retention techniques have been applied in image editing~\cite{p2p,cao2023masactrl}, the exploration of how attention mechanisms function in music editing remains unexplored.

To address these issues, we conduct an in-depth probing analysis on different attention maps within AudioLDM 2~\cite{audioldm2}, a diffusion-based model commonly used as the backbone for existing music editing methods. We reveal a key insight: cross-attention maps encompass details regarding distinct musical characteristics, and interventions on these maps result in ineffective modifications. In contrast, self-attention maps are essential for preserving the temporal structure of the source music during its conversion into the target music. This finding suggests that manipulating self-attention maps may yield better editing results. Building on this insight, we introduce Melodia, a training-free and source-prompt-free approach that selectively manipulates self-attention maps in specific layers during the diffusion model's denoising process. The source-prompt-free nature lowers the usage barrier for non-expert users. By building an attention repository to store self-attention information from the source music and applying it during editing, our method achieves a good balance between textual adherence and structural integrity without requiring any textual description of the source music. Fig.~\ref{fig:first} demonstrates that our proposed method Melodia achieves superior structural preservation compared to existing approaches. Furthermore, to address the challenge of evaluating the balance between textual adherence and structural integrity, we propose two novel metrics, Adherence-Structure Balance Score (ASB) and Adherence-Musicality Balance Score (AMB) that comprehensively measure this balance in music editing tasks.
Additionally, we construct MelodiaEdit, a music editing benchmark covering diverse editing scenarios.

In summary, our main contributions are as follows:
\begin{itemize}
\item We conduct a thorough analysis of attention maps within the diffusion-based model, revealing the unique functions of cross-attention and self-attention maps in music editing, with self-attention playing a critical role in preserving temporal structure.

\item We propose two novel evaluation metrics, ASB and AMB, to assess the balance between textual adherence and structural integrity, and construct MelodiaEdit benchmark covering diverse editing scenarios.

\item We propose Melodia, a novel training-free music editing technique that selectively manipulates self-attention maps during denoising, achieving an optimal balance between textual adherence and structural integrity without requiring textual descriptions of the source music.

\item Through comprehensive subjective and objective evaluations across three datasets, our approach consistently outperforms existing methods for intra-stem music editing.
\end{itemize}

\section{Related Work}

\subsection{Text-to-music generation}
Text-to-music generation follows two main paradigms: autoregressive (AR) models and diffusion-based models. AR models like MusicLM~\cite{agostinelli2023musiclm} introduced a music-text embedding space, while MusicGen~\cite{musicgen} improved controllability through text and melody guidance. Diffusion-based models transform noise into musical structures through iterative denoising. AudioLDM~\cite{audioldm} applied latent diffusion to text-conditioned generation, while Mousai~\cite{mousai} and MusicLDM~\cite{chen2024musicldm} further advanced quality and length capabilities. Recently, AudioLDM 2~\cite{audioldm2} improved musical structure representation, MelodyFlow~\cite{melodyflow} advances the field with flow matching techniques, and Stable Audio Open~\cite{evans2025stableaudioopen} scales diffusion transformers to generate extended stereo music. As these models have made significant progress in music generation, researchers have expanded their focus to music editing with its unique challenges.

\subsection{Text-to-music editing}

\textbf{Specialized Training Methods}. MusicLM~\cite{agostinelli2023musiclm} utilizes MuLan~\cite{huang2022mulan} embedding space for style editing, while MusicGen~\cite{musicgen} facilitates editing by conditioning generation on an original audio's chromagram with text prompts for desired changes. These two models offer limited editing capabilities as a secondary function. And models like AUDIT~\cite{wang2023audit} and InstructME~\cite{han2023instructme} train diffusion models specifically for inter-stem editing. While effective, these methods require extensive training on text-audio pairs.

\textbf{Fine-tuning Approaches} adapt pre-trained models through targeted optimization.~\citet{plitsis2024investigating} demonstrate techniques adapted from image editing, such as DreamBooth~\cite{ruiz2023dreambooth} for audio personalization. Instruct-MusicGen~\cite{zhang2024instructmusicgen} exemplifies this category by fine-tuning the pre-trained MusicGen model with an instruction-following strategy. While requiring less training data, these methods remain computationally intensive. 

\textbf{Zero-shot Editing Methods} manipulate music without additional training. MusicMagus~\cite{magus} uses cross-attention constraint and word swapping but requires specific keywords describing the original music to find editing directions. MEDIC~\cite{liu2024medic} unifies mutual self-attention control and cross-attention control. Yet, these methods focus on manipulating attention mechanisms without providing interpretability insights into how these mechanisms function in music diffusion models. Additionally, existing methods including DDPM Friendly~\cite{manor2024ddpm} and SDEdit~\cite{meng2021sdedit} 
lack explicit structure guidance from the original music, making it difficult to preserve temporal structure during editing.

\section{METHOD}
\subsection{Preliminary}
\textbf{Latent Diffusion Model (LDM)}~\cite{rombach2022high} is a form of diffusion model trained within low-dimensional latent space. Given data $x$, LDM employs a variational autoencoder (VAE)~\cite{kingma2013auto} encoder $\mathcal{E}$ to compress $x$ into the latent $z$ and a corresponding decoder $\mathcal{D}$ to reconstruct the data. In the diffusion process, the initial latent $z_0$ is converted to a sample $z_T$ by adding random noise $\epsilon_t\sim\mathcal{N}(0,1)$ with $T$ iterations. The denoising process recovers $z_0$ from $z_T$ utilizing a trained denoiser $\epsilon_\theta$.

\textbf{Cross-Attention (CA) Mechanism}, which establishes connections between inputs of different modalities, is crucial for implementing a conditional denoiser $\epsilon_\theta(z_t,t,y)$ conditioned by $y$. Output $\phi^{\rm c}$ of a CA layer is defined as:
\begin{align}
	\phi^{\rm c} = \text{Cross-Attent}&\text{ion}(Q^{\rm c},K^{\rm c},V^{\rm c}) = M^{\rm c}\cdot V^{\rm c} \\
	M^{\rm c}=&{\rm Softmax}\Big(\frac{Q^{\rm c}{K^{\rm c}}^\top}{\sqrt{d^{\rm c}}}\Big) \\
	Q^{\rm c}=W_{Q^{\rm c}}\cdot \varphi(z_t),\ \ K^{\rm c}&=W_{K^{\rm c}}\cdot \tau(y),\ V^{\rm c}=W_{V^{\rm c}}\cdot \tau(y)
\end{align}
where $\varphi(z_t)\in\mathbb{R}^{N\times d_\epsilon}$ denotes a flattened intermediate representation of hidden spatial features in the denoiser $\epsilon_\theta$, and  $\tau(\cdot)$ is a modality-specific encoder introduced to embed inputs $y$ of various modalities into unified embeddings $\tau(y)\in\mathbb{R}^{M\times d_\tau}$. $W_{Q^{\rm c}}\in\mathbb{R}^{d^{\rm c}\times d^\epsilon}$, $W_{K^{\rm c}},\, W_{V^{\rm c}}\in\mathbb{R}^{d^{\rm c}\times d^\tau}$ are learnable projection matrices. With $y$, conditional LDM is enabled to generate desired data with a conditional denoiser $\epsilon_\theta(z_t, t, y)$. Moreover, adopted from Ho et al.~\cite{ho2022classifier}, conditional LDM uses Classifier-Free Guidance strength (CFG) to control the strength of conditions.

\textbf{Self-Attention (SA) Mechanism} focuses on information processing within the latent itself by establishing connections among spatial patches. Output $\phi^{\rm s}$ of SA is defined as:
\begin{align}
	\phi^{\rm s} = \text{Self-Attent}&\text{ion}(Q^{\rm s},K^{\rm s},V^{\rm s}) = M^{\rm s}\cdot V^{\rm s} \\
	M^{\rm s}=&{\rm Softmax}\Big(\frac{Q^{\rm s}{K^{\rm s}}^\top}{\sqrt{d^{\rm s}}}\Big) \\
	Q^{\rm s}=W_{Q^{\rm s}}\cdot \varphi(z_t),\ \ K^{\rm s}&=W_{K^{\rm s}}\cdot \varphi(z_t),\ V^{\rm s}=W_{V^{\rm s}}\cdot \varphi(z_t)
\end{align}
Unlike CA, query $Q^{\rm s}$, key $K^{\rm s}$ and value $V^{\rm s}$ of SA are all derived from the hidden spatial feature $\varphi(z_t)$. Therefore, it can capture dependencies within patches of the input latent.

\textbf{AudioLDM 2}~\cite{audioldm2}, an open-source conditional LDM-based audio generation model, is the foundation model of our work.It contains 16 layers with cross-attention and self-attention mechanisms in its UNet architecture.
% \textbf{Detailed introduction in the Appendix}.
% In AudioLDM 2, the conditional denoiser $\epsilon_\theta(z_t,t,y)$ is implemented by a 16-layer Transformer UNet (T-UNet)~\cite{ronneberger2015unet, vaswani2017transformer}, where each layer consists of several transformer blocks. The first self transformer block is a stack of self-attention layers and feedforward networks~\cite{audioldm2}. To incorporate the text condition information $y$ related to the desired music clip, the last two transformer blocks receive condition embeddings $\tau_\text{GPT-2}(y)$ and $\tau_\text{T5}(y)$, respectively. $\tau_\text{T5}(y)$ is encoded through the FLAN-T5~\cite{chung2024t5} language model, while $\tau_\text{GPT-2}(y)$ is encoded by both the FLAN-T5 language model and the CLAP~\cite{elizalde2023clap} text encoder, and then fed into the GPT-2~\cite{radford2019gpt} model to approximate AudioMAE~\cite{huang2022masked} based acoustic features. The selection of AudioLDM 2 as the backbone is motivated by its capability for robust text-guided music generation, which shows immense potential for application in zero-shot text-guided music editing.

\begin{table}
  \caption{Probing accuracy of CA map in different layers.}
  \label{tab:ca-analysis}
  \small
  \setlength{\tabcolsep}{2pt}  % 减小列间距，默认是6pt
  \resizebox{\linewidth}{!}{
    \begin{tabular}{c|cccccc|ccc}
    \toprule
     Class & Layer 1  & Layer 4 & Layer 6 & Layer 10 & Layer 13 & Layer 16 & Avg. \\
    \midrule
    piano & 0.90 & 1.00  & 0.50 & 0.97 & 0.97 & 0.95 & 0.88 \\
    accordion & 0.80 & 0.77 & 1.00 & 0.90 & 0.70 & 0.72 & 0.82 \\
    \midrule
    jazz & 0.97 & 0.72 & 0.47 & 0.72 & 1.00 & 0.85 & 0.79 \\
    country & 0.90 & 0.70 & 0.32 & 0.82 & 1.00 & 0.85 & 0.77 \\
    \midrule
    sad & 0.95 & 0.90 & 0.57 & 0.95 & 0.77 & 0.82 & 0.83 \\
    happy & 0.70 & 0.60 & 0.62 & 0.90 & 0.77 & 0.60 & 0.70 \\
    
    \bottomrule
    \end{tabular}
  }
\end{table}

\begin{table}
  \caption{Probing accuracy of SA map in different layers.}
  \label{tab:sa-analysis}
  \small
  \setlength{\tabcolsep}{2pt}  % 减小列间距，默认是6pt
  \resizebox{\linewidth}{!}{
      \begin{tabular}{c|cccccc|ccc}
        \toprule
         Class & Layer 1  & Layer 4 & Layer 6 & Layer 10 & Layer 13 & Layer 16 & Avg. \\
        \midrule
        piano & 0.15 & 0.37 & 0.12 & 0.25 & 0.70 & 0.02 & 0.27 \\
        accordion & 0.22 & 0.05 & 0.17 & 0.45 & 0.65 & 0.40 & 0.32 \\
        \midrule
        jazz & 0.32 & 0.17 & 0.05 & 0.37 & 0.22 & 0.07 & 0.20 \\
        country & 0.10 & 0.02 & 0.05 & 0.05 & 0.67 & 0.10 & 0.17 \\
        \midrule
        sad & 0.02 & 0.15 & 0.27 & 0.27 & 0.67 & 0.60 & 0.33 \\
        happy & 0.05 & 0.65 & 0.05 & 0.27 & 0.30 & 0.17 & 0.38 \\
    
        \bottomrule
      \end{tabular}
    }
\end{table}

\begin{figure}
    \centering
    \includegraphics[width=0.9\linewidth]{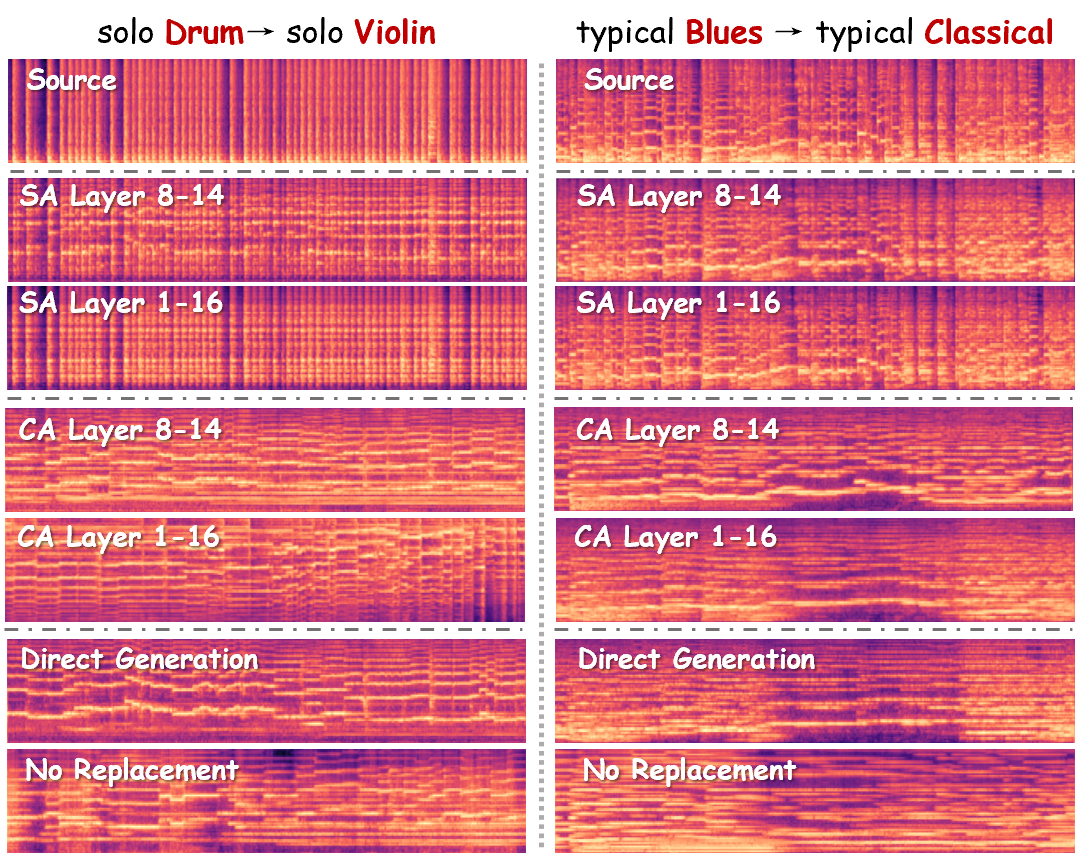}
    \caption{Results of cross-attention map and self-attention map replacement in different layers of the AudioLDM 2.}
    \label{fig:sa-ca}
\end{figure}

\subsection{Probing Analysis on Attention Maps}
\label{section analysis}
In this section, we analyze how cross-attention and self-attention maps in AudioLDM2~\cite{audioldm2} contribute to music editing effectiveness. Understanding these mechanisms is critical for developing methods that can transform musical attributes while preserving the original structure.

\textbf{Probing Methodology}. In AudioLDM 2~\cite{audioldm2}, attention mechanisms are the key to its successful music generation. However, \textbf{it remains unclear whether these attention maps are merely weight matrices or contain rich semantic representations of music}. To address this, inspired by probing analysis methods in NLP~\cite{clark2019Bertattention,liu2019linguistic}, we build datasets and train classification networks to explore attention map properties. The principle is straightforward: If a classifier can accurately categorize attention maps into different classes, these maps must encode meaningful semantic information beyond mere weighting. 
We construct three prompt datasets targeting fundamental musical dimensions: instruments (16 types, \emph{a solo [instrument] music}), styles (11 types, \emph{a typical [style] music}), and moods (8 types, \emph{a [mood] music}). 
For each prompt, we extract cross-attention maps corresponding to specific keywords ([instrument], [style], and [mood]) and self-attention maps from different layers during generation. 
Finally, we train a simple MLP classifier to determine if these maps encode specific musical attributes. High classification accuracy indicates that attention maps encode substantial information about musical attributes rather than functioning merely as weighting mechanisms.
% \textbf{For details about our probing method and prompt datasets, please refer to Appendix}.

\begin{figure*}
  \centering
  \includegraphics[width=0.9\textwidth]{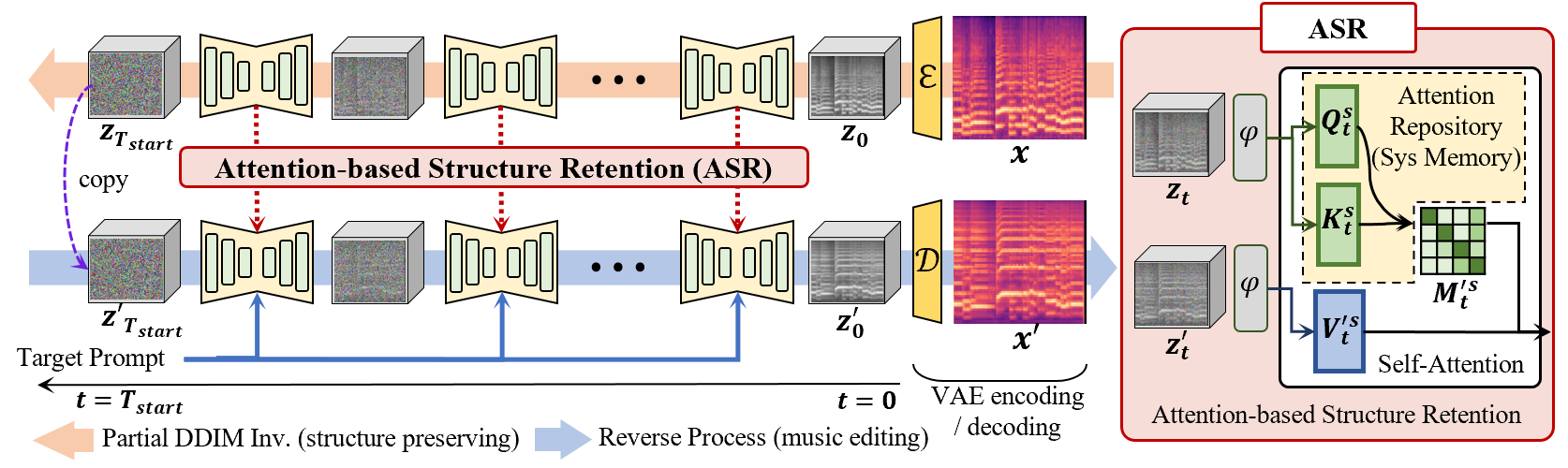}
  \caption{(Left) Overview of Melodia. 
  %We first employs a VAE~\cite{kingma2013auto} encoder $\mathcal{E}$ to compress original music $x$ to the latent space as $z_0$. Then, we invert $z_0$ into $z_{T_\text{start}}$ via DDIM Inversion~\cite{song2020ddim} within AudioLDM 2~\cite{audioldm2}, where $T_\text{start}\leq T$ is the specific timestep where we start the reverse process. While performing the inversion process, we store the self-attention queries $Q^{\rm s}_t$ and keys $K^{\rm s}_t$ corresponding to the latent $z_t$ at each inversion step $t$. This information stored in the attention repository will provide accurate guidance for structure preserving for the subsequent music editing process via Attention-based Structure Rention (ASR). Finally, a VAE decoder will be used to decode the edited latent $z'_0$ back to the corresponding mel spectrogram.
  % 
  $\quad$(Right) Detailed explanation of Attention-based Structure Retention (ASR). 
  %We manipulate the information in self-attention layer at each timestep $t$ of the reverse process. Specially, the query ${Q'}_t^{\rm s}$ and key ${K'}_t^{\rm s}$ are substituted with the stored query $Q_t^s$ and key $K_t^{\rm s}$ from identical timestep $t$. The corresponding spatial attention map $M'^{\rm s}_t$ is thus substituted with $M^{\rm s}_t$, which is related to the original latent $z_0$, providing necessary temporal structure information of original music for precise text-guided intra-stem music editing. 
  %
  }
  \label{framework}
  
\end{figure*}

\textbf{Probing Results on Cross-Attention Maps}. Tab.~\ref{tab:ca-analysis} shows our classification results on cross-attention maps across different layers. Our classifier achieves high accuracy across all three classification tasks, with average accuracy rates exceeding 70\%, indicating that cross-attention maps contain rich semantic information about musical characteristics rather than simple weighting.
This explains why direct manipulation of cross-attention maps causes unsuccessful music editing outcomes.
Fig.~\ref{fig:sa-ca} shows that when replacing cross-attention maps (layers 8-14 or 1-16), the temporal structure of the edited music significantly differs from the source music, yet aligns closely with the music directly generated from the target prompt.
% For example, in the left part of the Fig.~\ref{fig:sa-ca}, the source drum music shows distinct rapid rhythmic patterns in spectrograms, but cross-attention manipulation completely loses these features, producing violin music that resembles direct generation rather than successful editing. 
% \textbf{For complete experimental results, please refer to the Appendix}.

\textbf{Probing Results on Self-Attention Maps}. Tab.~\ref{tab:sa-analysis} shows that our classifier struggles with self-attention maps, achieving low average accuracy ($<$40\%) significantly below cross-attention results.
This suggests self-attention maps do not encode categorical information about musical attributes. Instead, our further analysis shows they capture temporal features critical for music coherence, such as melody and rhythm.
This finding parallels observations in image diffusion models, where self-attention maps preserve the spatial structure of images~\cite{liu2024towardsunderstand}.
To validate this, we conducted self-attention map replacement experiments. 
Fig.~\ref{fig:sa-ca} shows that replacing self-attention maps in layers 8-14 successfully changes attributes while preserving original melody and rhythm—the edited violin music maintains the source drum's beat pattern.
Without replacement, the music loses source temporal structure and resembles direct generation from the target prompt.
However, replacing all layers (1-16) partially preserves original timbre instead of complete transformation to violin.
These experimental results support our idea that self-attention maps are essential for preserving temporal structure during music editing, with selective layer replacement showing promising results.

\subsection{Overview of Our Approach}

Building on our exploration of attention layers, the fundamental innovation of our approach lies in preserving the temporal structure of the original music clip $x_0$ within self-attention map manipulation. Based on this, we propose a straightforward yet effective approach named Melodia. 

As shown in Fig.~\ref{framework}, in our approach, we first obtain the latent $z_0$ of the original music clip via a VAE~\cite{kingma2013auto} encoder $\mathcal{E}$, and then collect the SA features of $z_0$ during the \textit{Partial DDIM Inversion}~\cite{song2020ddim, manor2024ddpm} process (hereafter DDIM inversion), a version that applies the inversion only up to $\ T_{\rm start}\in [0,T], T=1000$~\cite{manor2024ddpm}. Specifically, for predefined time steps $t\in\{0,\dots,T_{\rm start}\}$, we invert the original latent $z_0$ into an editable intermediate latent $z_{T_{\rm start}}$ and store self-attention queries $Q^{\rm s}_t$ and keys $K^{\rm s}_t$ at each time step to build an attention repository in this process. The repository storing self-attention features of the original music clip is used to provide guidance for subsequent music editing. In the music editing process, we transform the features stored in the attention repository into original music clip structural guidance through the proposed Attention-based Structure Retention (ASR) method, which is based on manipulation of the self-attention mechanisms of pre-trained diffusion models, without extra fine-tuning or new blocks. Similarly to the guidance of the given target prompt, this guidance also modulates the denoising process. Finally, the output $z'_0$ is decoded into a readable mel spectrogram of the desired music clip via the corresponding VAE decoder $\mathcal{D}$.
% \textbf{See the pseudocode of our algorithm in the Appendix}.

\begin{figure}
    \centering
    \includegraphics[width=0.9\linewidth]{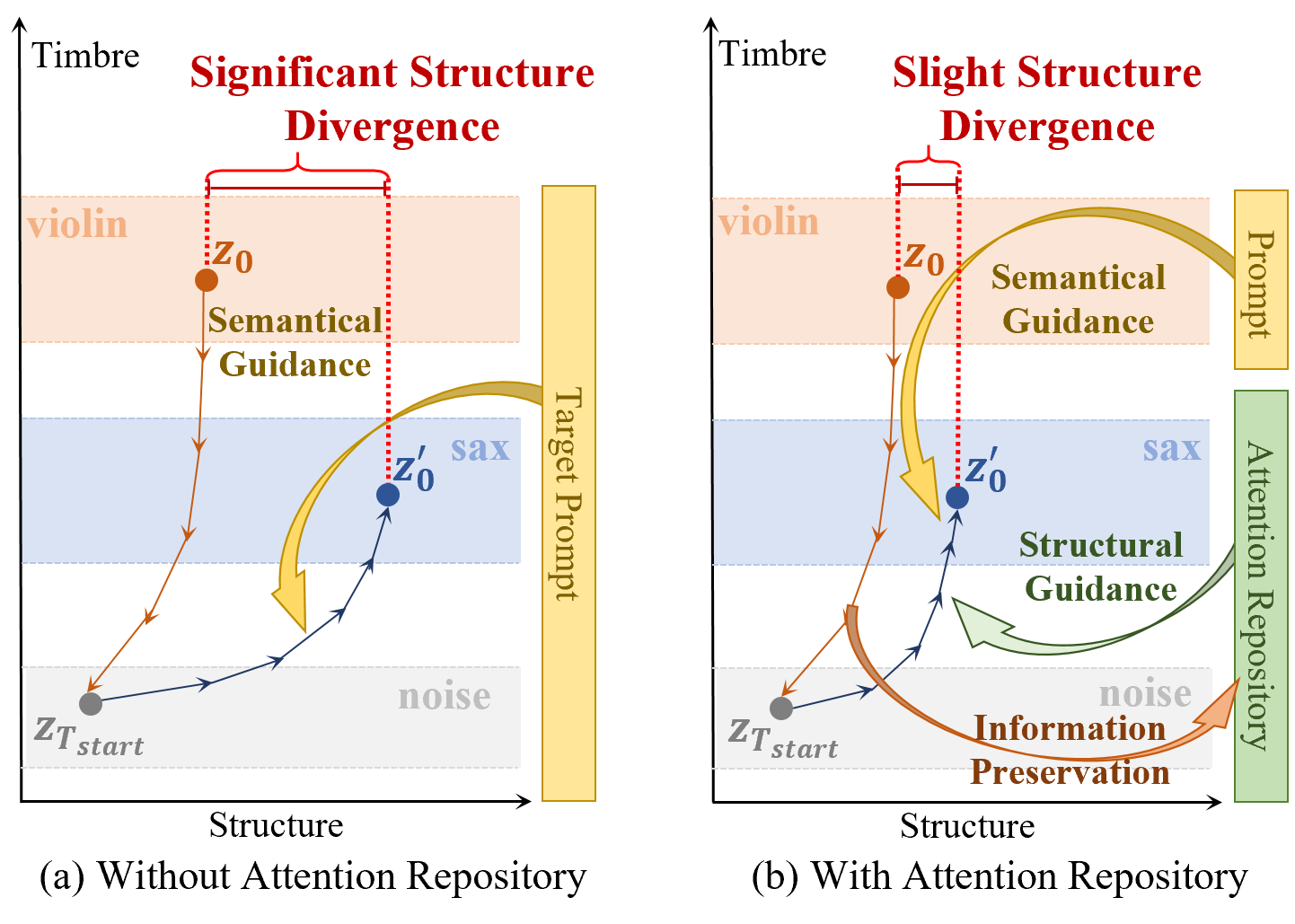}
    \caption{Intuitive Illustration of DDIM Inversion and Reverse Process with Attention Repository based Structure Guidance. The orange and blue paths respectively refer to DDIM Inversion path and reverse path.}
    \label{fig:grad}
\end{figure} 

\subsection{DDIM Inversion with Attention Repository}
\label{sect:ddim}
Adopted from Content-Style Modeling in Multi-Domain Analysis~\cite{sorensen2021linearsub, shrestha2024cssub}, we assume that any music sample $x^{(n)}$  from domain $n$ can be represented as a bijective function $\mathbf{g}$ of content $\mathbf{c}$ and style $\mathbf{s^{(n)}}$ (i.e. timbre) components:
\begin{equation}
    \mathbf{c}\sim \mathbb{P}_\mathbf{c},\ \mathbf{s}^{(n)}\sim\mathbb{P}_{\mathbf{s}^{(n)}},\ x^{(n)}=\mathbf{g}(\mathbf{c},\mathbf{s}^{(n)})
\end{equation}
where $\mathbb{P}_\mathbf{s^{(n)}}$ and $\mathbb{P}_\mathbf{c}$ are distributions of the style components in $n$th domain and the content components respectively. Assuming that the VAE encoder $\mathcal{E}$ is a bijective function and its inverse is the VAE decoder $\mathcal{D}$, the combination of two bijective functions $\mathbf{g}$ and $\mathcal{E}$ ensures that each $z=\mathcal{E}(x)=\mathcal{E}(\mathbf{g}(\mathbf{c}, \mathbf{s}))$ corresponds to a unique style and content component pair $(\mathbf{c},\mathbf{s})$. 

Based on the above assumption, an intuitive illustration is presented in Fig.~\ref{fig:grad}, showing a Structure-Timbre sampling space of the music latent diffusion model~\cite{audioldm, audioldm2}. In our music editing process, the starting latent $z'_{T_{\rm start}}$, copied from DDIM Inversion~\cite{song2020ddim} prior $z_{T_{\rm start}}$, provides implicit structure guidance for generation. However, as shown in Fig.~\ref{fig:grad} (a), relying on this implicit guidance leads to significant structure divergence in the editing process~\cite{song2020ddim, mou2023dragon, tumanyan2023plug}. Since the target prompt $y$ contains rich semantics of music, it can provide strong semantic guidance for editing, which is significantly stronger than the implicit structure guidance. Therefore, we build an attention repository to store SA features at each inversion time step in the system memory, and use the stored features to provide explicit structure guidance for the editing process, reducing divergence (shown in Fig.~\ref{fig:grad} (b)).

% \textbf{See details of GPU memory and inference time overhead of attention repository in the Appendix}.
\subsection{Attention-based Structure Retention} Attention-based Structure Retention (ASR) is the key to transforming the stored attention features into structure guidance. As shown in the right part of Fig.~\ref{framework}, to achieve this, the SA map $M'^{\rm s}_t$ of the target latent $z'_t$ is derived from the stored original SA queries $Q^{\rm s}_t$ and keys $K^{\rm s}_t$ at each reverse time step $t\in \{0,\dots,T_{\rm start}\}$ instead of attention features of $z'_t$. The manipulated SA can defined as:
\begin{align}
	\phi'^{\rm s}_t = M'^{\rm s}_t& \cdot V'^s_t \\
	M'^{\rm s}_t = {\rm Softm}&{\rm ax} \Big( \frac{Q^{\rm s}_t{K^{\rm s}_t}^\top}{\sqrt{d^s}} \Big) \\
	Q^{\rm s}_t=W_{Q^{\rm s}}\cdot \varphi(z_t),\ \ K^{\rm s}_t=W_{K^{\rm s}}&\cdot \varphi(z_t),\ {V'}^{\rm s}_t=W_{V^{\rm s}}\cdot \varphi(z'_t)
\end{align}
where $\phi'^{\rm s}_t$ is the output of a manipulated self-attention layer and $V'^{\rm s}_t$ is the projected self-attention values of target latent $z'_t$ at the specific time step $t$. In addition, we apply this manipulation to layers 8-14 of AudioLDM 2~\cite{audioldm2}. Layer selection is analyzed in the Experiments section.

\begin{table*}[htbp]
\centering
\normalsize
\setlength{\tabcolsep}{0.1em}
\caption{Objective evaluation results across three datasets.}
\label{tab:objective_comparison}
\begin{threeparttable}
\resizebox{\linewidth}{!}{
\begin{tabular}{l|*{6}{c}|*{6}{c}|*{6}{c}}
\toprule
& \multicolumn{6}{c|}{\textbf{MusicDelta}} & \multicolumn{6}{c|}{\textbf{ZoME-Bench}} & \multicolumn{6}{c}{\textbf{MelodiaEdit}} \\
\cmidrule(lr){2-7} \cmidrule(lr){8-13} \cmidrule(lr){14-19}
\textbf{Method} & \textbf{CLAP}$\uparrow$ & \textbf{LPAPS}$\downarrow$ & \textbf{Chroma}$\uparrow$ & \textbf{FAD}$\downarrow$ & \textbf{ASB}$\uparrow$ & \textbf{AMB}$\uparrow$ & \textbf{CLAP}$\uparrow$ & \textbf{LPAPS}$\downarrow$ & \textbf{Chroma}$\uparrow$ & \textbf{FAD}$\downarrow$ & \textbf{ASB}$\uparrow$ & \textbf{AMB}$\uparrow$ & \textbf{CLAP}$\uparrow$ & \textbf{LPAPS}$\downarrow$ & \textbf{Chroma}$\uparrow$ & \textbf{FAD}$\downarrow$ & \textbf{ASB}$\uparrow$ & \textbf{AMB}$\uparrow$ \\
\midrule
SDEdit & 0.17 & 6.82 & 0.20 & 1.47 & 0.00 & 0.00 & 0.12 & 6.85 & 0.19 & 1.27 & 0.00 & 0.00 & 0.34 & 5.46 & 0.49 & 1.15 & 0.29 & 0.29 \\
MusicGen & 0.22 & 6.27 & 0.24 & 0.98 & 0.24 & 0.31 & \textbf{0.29} & 6.42 & 0.22 & 0.92 & 0.25 & 0.46 & \underline{0.36} & 5.11 & 0.59 & 1.00 & 0.25 & 0.54 \\
MusicMagus & 0.20 & \underline{5.06} & \underline{0.29} & 0.95 & 0.28 & 0.29 & 0.22 & \underline{4.82} & 0.26 & 0.72 & \underline{0.63} & 0.64 & 0.27 & \underline{3.32} & \textbf{0.73} & \textbf{0.57} & 0.00 & 0.00 \\
DDPM-Friendly & \textbf{0.35} & 5.66 & 0.27 & \underline{0.88} & \underline{0.58} & \underline{0.74} & \underline{0.23} & 5.70 & \underline{0.27} & \underline{0.68} & 0.49 & \underline{0.72} & 0.34 & 4.06 & \underline{0.70} & 0.67 & \underline{0.59} & \underline{0.70} \\
DDIM Inversion & 0.30 & 5.93 & 0.26 & 1.03 & 0.45 & 0.60 & 0.22 & 5.82 & 0.25 & 0.77 & 0.44 & 0.59 & 0.35 & 4.58 & 0.65 & 0.90 & 0.48 & 0.67 \\
\midrule
\textbf{Melodia (Ours)} & \underline{0.34} & \textbf{4.01} & \textbf{0.32} & \textbf{0.56} & \textbf{1.00} & \textbf{1.00} & \textbf{0.29} & \textbf{3.90} & \textbf{0.29} & \textbf{0.47} & \textbf{1.00} & \textbf{1.00} & \textbf{0.39} & \textbf{3.11} & 0.68 & \underline{0.65} & \textbf{1.00} & \textbf{0.88} \\
\bottomrule
\end{tabular}
}
\begin{tablenotes}
\footnotesize
\item 
\end{tablenotes}
\end{threeparttable}
\end{table*}

\section{Experiments}

\subsection{Baselines}
\label{sec:base}
To comprehensively evaluate our approach, we conduct comparisons with several state-of-the-art music editing approaches, including DDPM-Friendly~\cite{manor2024ddpm}, MusicMagus~\cite{magus}, SDEdit~\cite{meng2021sdedit}, DDIM Inversion~\cite{song2020ddim} and MusicGen~\cite{musicgen}, for both our objective and subjective evaluations. For the music generation model MusicGen~\cite{musicgen}, we utilize its melody-conditioned medium checkpoint to perform editing tasks. We do not include AUDIT~\cite{wang2023audit}, InstructMe~\cite{han2023instructme} in our comparisons, as these methods were developed for inter-stem music editing and their implementations are not publicly available. Similarly, we could not compare with MEDIC~\cite{liu2024medic} as its source code has not been released.

\subsection{Metrics}
\label{sec:metirc}
For objective evaluation, we evaluate the results using four standard metrics. CLAP~\cite{wu2023largeclap} measures the adherence of the result to the target prompt (higher is better). LPAPS~\cite{paissan2023audiolpaps} measures the preservation of temporal structure and coherence between the edited audio and source audio (lower is better), while Chroma~\cite{musicgen} quantifies the preservation of harmonic, melodic, and pitch elements (higher is better). FAD~\cite{kilgour2018fad} measures the distributional
difference between source and edited music (lower is better). 

However, individual metrics can be misleading when evaluating editing performance. Methods that barely modify the source music achieve high structure-related scores but low CLAP scores, creating a false impression of success. To address this, we propose two composite metrics: Adherence-Structure Balance Score (ASB) and Adherence-Musicality Balance Score (AMB). We use harmonic mean (like F1-score) to ensure neither text adherence nor structure preservation dominates the assessment, thereby achieving balance in music editing evaluation.

For subjective evaluation, we introduce the Music Editing Balance (MEB) metric to capture human perception. Additionally, adapted from MusicMagus~\cite{magus}, we employ Relevance (REL) to assess how well the edited music aligns with the target prompt, and Structural Consistency (CON) to evaluate the consistency of the pitch contour and structural aspects. These metrics are rated on a five-point Likert scale~\cite{likert1932technique}, with higher scores being better.

% \textbf{More details of these metrics can be found in Appendix}. 

\begin{figure}
    \centering
    \includegraphics[width=0.9\linewidth]{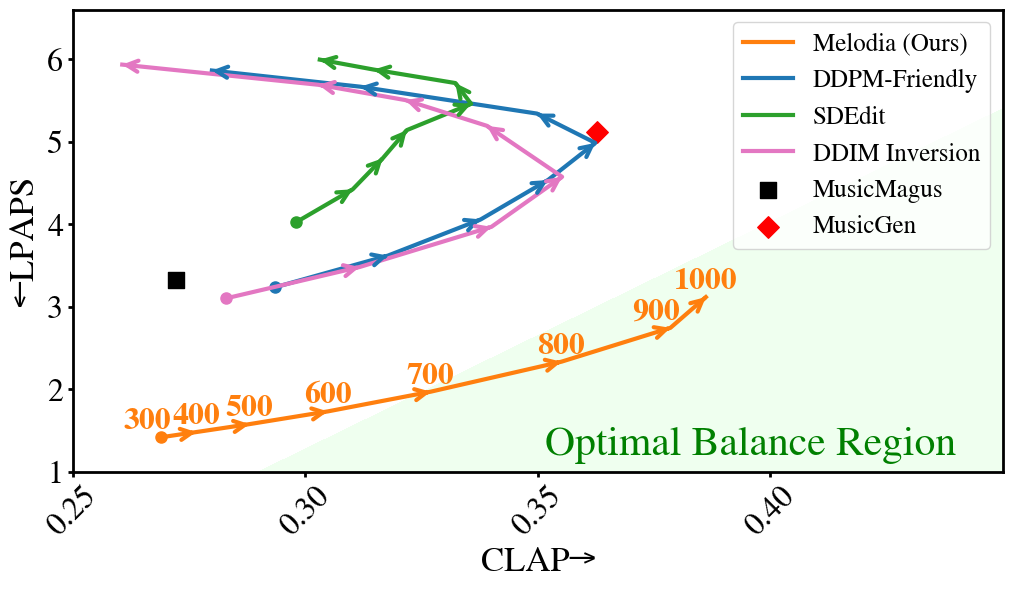}
    \caption{Quantitative Comparison with methods over $T_\text{start}$ range of 300-1000 on MelodiaEdit. The highlighted region is the optimal balance region where shows both text adherence and structural integrity. Our method outperforms other approaches across all $T_\text{start}$ values.}
    \label{fig:clap-lpaps}
\end{figure}

\textbf{Adherence-Structure Balance Score (ASB)} evaluates the balance between the adherence of the edited music to the target prompt and structural preservation to the source music. ASB integrates CLAP and LPAPS:
\begin{equation}
\text{ASB} = \frac{2 \cdot \mathcal{S}(\mathcal{N}(s_\text{CLAP})) \cdot \mathcal{S}(\mathcal{N}(-s_\text{LPAPS}))}{\mathcal{S}(\mathcal{N}(s_\text{CLAP})) + \mathcal{S}(\mathcal{N}(-s_\text{LPAPS}))}
\end{equation}
where $s_\text{CLAP}$ is the CLAP score, $s_\text{LPAPS}$ is the LPAPS score. We use the negative of $s_\text{LPAPS}$ because lower $s_\text{LPAPS}$ values indicate better preservation.

We apply Z-score normalization $\mathcal{N}(\cdot)$ and Min-Max scaling $\mathcal{S}(\cdot)$ to ensure scores fall within $[0,1]$, with 1 indicating best performance and 0 the worst across evaluated methods.

% \textbf{Detailed procedures are provided in the Appendix.}

\textbf{Adherence-Musicality Balance Score (AMB)} assesses  the balance between the adherence of the edited music to the target
prompt and musicality preservation of the source music. AMB combines CLAP with Chroma:
\begin{equation}
\text{AMB} = \frac{2 \cdot \mathcal{S}(\mathcal{N}(s_\text{CLAP})) \cdot \mathcal{S}(\mathcal{N}(s_\text{Chroma}))}{\mathcal{S}(\mathcal{N}(s_\text{CLAP})) + \mathcal{S}(\mathcal{N}(s_\text{Chroma}))}
\end{equation}
where $s_\text{Chroma}$ is the Chroma similarity. Both components undergo the same two-step normalization process as in ASB.

\textbf{Music Editing Balance (MEB)} is our newly proposed perceptual metric that evaluates how well edited music balances adherence to the target prompt while preserving the original music's temporal structure and musicality. Higher MEB scores indicate better balance.

\subsection{Dataset}
\label{exp:dataset}
We evaluate our method using MusicDelta~\cite{bittner2014medleydb}, ZoME-Bench~\cite{liu2024medic}, and MelodiaEdit, our newly constructed benchmark. MelodiaEdit comprises 2,015 music-prompt pairs designed for comprehensive intra-stem editing evaluation. The dataset contains five subsets: three synthesized subsets with 1,090 pairs generated via AudioLDM 2~\cite{audioldm2}, and two authentic subsets with 925 pairs reconstructed from Medley-solos-DB~\cite{lostanlen2016realins} and GTZAN datasets.

\subsection{Objective Evaluation}
\label{section quantitative}

We evaluate our proposed method through a comparison with five competing approaches. All diffusion-based methods are implemented using the pre-trained AudioLDM 2~\cite{audioldm2} model with 200 inference steps. The target CFG of our method is set to 5.5.

% \textbf{Details on CFG analysis and experimental settings can be found in Appendix}. 

\textbf{Quantitative Comparison with baselines on datasets}. Table~\ref{tab:objective_comparison} compares our method with baseline approaches. Our method achieves competitive CLAP and the lowest LPAPS across all datasets, indicating superior text adherence and temporal structure preservation.
Melodia also maintains good Chroma and FAD performance, effectively preserving harmonic elements and audio quality.
While MusicMagus~\cite{magus} shows high Chroma and FAD on MelodiaEdit, this indicates editing failure as outputs remain unchanged from the source music, explaining its poor textual adherence.
The ASB and AMB results demonstrate the necessity of our proposed composite metrics. Several baselines achieve 0.00 on these metrics due to severe imbalances between text adherence and structure preservation, while Melodia achieves excellent performance on both composite metrics, demonstrating superior balance between editing effectiveness and structure preservation.

% \textbf{Task-specific results within each dataset are provided in the Appendix.}

\begin{table}[t]
\centering
\normalsize
\setlength{\tabcolsep}{0.1em}
\caption{Subjective evaluation results across three datasets.}
\label{tab:subjective_comparison}
\begin{threeparttable}
\resizebox{\linewidth}{!}{
\begin{tabular}{l|*{3}{c}|*{3}{c}|*{3}{c}}
\toprule
& \multicolumn{3}{c|}{\textbf{MusicDelta}} & \multicolumn{3}{c|}{\textbf{zoME-Bench}} & \multicolumn{3}{c}{\textbf{MelodiaEdit}} \\
\cmidrule(lr){2-4} \cmidrule(lr){5-7} \cmidrule(lr){8-10}
\textbf{Method} & \textbf{REL}$\uparrow$ & \textbf{CON}$\uparrow$ & \textbf{MEB}$\uparrow$ & \textbf{REL}$\uparrow$ & \textbf{CON}$\uparrow$ & \textbf{MEB}$\uparrow$ & \textbf{REL}$\uparrow$ & \textbf{CON}$\uparrow$ & \textbf{MEB}$\uparrow$ \\
\midrule
SDEdit & 1.24 & 1.18 & 1.15 & 1.73 & 1.35 & 1.78 & 2.32 & 2.58 & 2.52 \\
MusicGen & 2.43 & 1.92 & 2.18 & \textbf{3.02} & 1.59 & 2.04 & 2.87 & 1.75 & \underline{3.13} \\
MusicMagus & 1.95 & 2.67 & 2.31 & 1.88 & \underline{2.65} & 1.69 & 1.89 & \textbf{3.84} & 1.21 \\
DDPM-Friendly & \underline{3.09} & \underline{2.88} & \underline{3.02} & 2.54 & 2.01 & \underline{2.65} & \underline{2.58} & 2.92 & 2.78 \\
DDIM Inversion & 2.76 & 2.34 & 2.49 & 2.17 & 1.68 & 1.89 & 3.02 & 2.67 & 2.92 \\
\midrule
\textbf{Melodia (Ours)} & \textbf{3.21} & \textbf{3.59} & \textbf{3.46} & \underline{2.85} & \textbf{3.48} & \textbf{3.21} & \textbf{3.38} & \underline{3.65} & \textbf{3.81} \\
\bottomrule
\end{tabular}
}
\begin{tablenotes}
\footnotesize
 \item 
\end{tablenotes}
\end{threeparttable}
\end{table}

\begin{table*}[htbp]
\centering
\normalsize  % 使用稍大一点的字体
\setlength{\tabcolsep}{0.1em}  % 适中的列间距
\caption{Performance comparison across different layer selections for editing tasks.}
\label{tab:duoceng}
\begin{threeparttable}
\resizebox{\linewidth}{!}{
\begin{tabular}{l|*{6}{c}|*{6}{c}|*{6}{c}}
\toprule
& \multicolumn{6}{c|}{\textbf{Timbre Transfer}} & \multicolumn{6}{c|}{\textbf{Genre Transfer}} & \multicolumn{6}{c}{\textbf{Mood Transfer}} \\
\cmidrule(lr){2-7} \cmidrule(lr){8-13} \cmidrule(lr){14-19}
\textbf{Layers} & \textbf{CLAP}$\uparrow$ & \textbf{LPAPS}$\downarrow$ & \textbf{Chroma}$\uparrow$ & \textbf{FAD}$\downarrow$ & \textbf{ASB}$\uparrow$ & \textbf{AMB}$\uparrow$ & \textbf{CLAP}$\uparrow$ & \textbf{LPAPS}$\downarrow$ & \textbf{Chroma}$\uparrow$ & \textbf{FAD}$\downarrow$ & \textbf{ASB}$\uparrow$ & \textbf{AMB}$\uparrow$ & \textbf{CLAP}$\uparrow$ & \textbf{LPAPS}$\downarrow$ & \textbf{Chroma}$\uparrow$ & \textbf{FAD}$\downarrow$ & \textbf{ASB}$\uparrow$ & \textbf{AMB}$\uparrow$ \\
\midrule
None & 0.34 & 4.39 & 0.71 & 0.87 & 0.00 & 0.00 & \underline{0.39} & 4.64 & 0.76 & 0.87 & 0.00 & 0.41 & 0.31 & 4.49 & 0.73 & 0.80 & 0.00 & 0.18 \\
1-16 & 0.34 & \textbf{2.65} & \textbf{0.81} & \textbf{0.55} & 0.00 & 0.00 & 0.34 & \textbf{2.54} & \textbf{0.84} & \textbf{0.54} & 0.00 & 0.00 & 0.28 & \textbf{1.07} & \textbf{0.96} & \textbf{0.11} & 0.00 & 0.00 \\
6-16 & 0.35 & \underline{2.96} & \underline{0.80} & \underline{0.57} & 0.22 & 0.22 & 0.35 & \underline{2.59} & 0.81 & \textbf{0.54} & 0.29 & 0.27 & 0.28 & \underline{1.08} & \underline{0.95} & \textbf{0.11} & 0.00 & 0.00 \\
10-12 & 0.39 & 3.93 & 0.76 & 0.89 & 0.37 & 0.56 & \textbf{0.40} & 3.92 & 0.76 & 0.88 & 0.51 & 0.43 & \textbf{0.35} & 4.48 & 0.72 & 0.99 & 0.01 & 0.14 \\
1-7 & \underline{0.40} & 4.16 & 0.73 & 0.94 & 0.23 & 0.32 & \textbf{0.40} & 4.16 & 0.73 & 0.94 & 0.37 & 0.00 & 0.29 & 4.46 & 0.70 & 1.07 & 0.02 & 0.00 \\
\midrule
8-14 & \textbf{0.42} & 3.49 & 0.75 & 0.81 & \textbf{0.68} & \textbf{0.57} & \textbf{0.40} & 2.63 & \underline{0.82} & \underline{0.56} & \textbf{0.98} & \textbf{0.90} & \underline{0.34} & 2.61 & 0.79 & \underline{0.46} & \textbf{0.67} & \textbf{0.49} \\
\bottomrule
\end{tabular}
}
\begin{tablenotes}
\footnotesize
\item 
\end{tablenotes}
\end{threeparttable}
\end{table*}

\textbf{Quantitative Comparison with baselines across $T_\text{start}$s}. 
To quantitatively evaluate the adherence of edited music to the target prompt and the fidelity to the original music, we plot the CLAP-LPAPS results on MelodiaEdit for all methods in Fig.~\ref{fig:clap-lpaps}. MusicMagus~\cite{magus} is tested only at $T_\text{start}=1000$ following its original setting, while other diffusion-based methods are evaluated across multiple $T_\text{start}$ values ranging from 300-1000. The results show that MusicGen~\cite{musicgen} achieves good textual adherence but poor structural preservation. Notably, other diffusion-based methods also fail to precisely preserve the structure of the source music. In contrast, Melodia addresses this issue by introducing structure guidance and achieves better LPAPS scores for any level of CLAP.

% \begin{table}[t]
%   \centering
%   \caption{Subjective Metrics Comparison across Different Datasets.}
%   \label{tab:subjective_comparison}
%   \footnotesize
%   \setlength{\tabcolsep}{6pt}
%   \begin{tabular}{l|l|ccc}
%     \toprule
%      Dataset & Method & REL$\uparrow$ & CON$\uparrow$ & MEB$\uparrow$ \\
%     \midrule
%     \multirow{6}{*}{MusicDelta} 
%     & SDEdit & 1.24 & 1.18 & 1.15 \\
%     & MusicGen & 2.43 & 1.92 & 2.18 \\
%     & MusicMagus & 1.95 & 2.67 & 2.31 \\
%     & DDPM-Friendly & \underline{3.09} & \underline{2.88} & \underline{3.02} \\
%     & DDIM Inversion & 2.76 & 2.34 & 2.49 \\
%     \cmidrule{2-5}
%     & \textbf{Melodia (Ours)} & \textbf{3.21} & \textbf{3.59} & \textbf{3.46} \\
%     \midrule
%     \multirow{6}{*}{zoME-Bench} 
%     & SDEdit & 1.73 & 1.35 & 1.78 \\
%     & MusicGen & \textbf{3.02} & 1.59 & 2.04 \\
%     & MusicMagus & 1.88 & \underline{2.65} & 1.69 \\
%     & DDPM-Friendly & 2.54 & 2.01 & \underline{2.65} \\
%     & DDIM Inversion & 2.17 & 1.68 & 1.89 \\
%     \cmidrule{2-5}
%     & \textbf{Melodia (Ours)} & \underline{2.85} & \textbf{3.48} & \textbf{3.21} \\
%     \midrule
%     \multirow{6}{*}{\begin{tabular}[c]{@{}l@{}}Supplementary\\dataset\end{tabular}} 
%     & SDEdit & 2.32 & 2.58 & 2.52 \\
%     & MusicGen & 2.87 & 1.75 & \underline{3.13} \\
%     & MusicMagus & 1.89 & \textbf{3.84} & 1.21 \\
%     & DDPM-Friendly & \underline{2.58} & 2.92 & 2.78 \\
%     & DDIM Inversion & 3.02 & 2.67 & 2.92 \\
%     \cmidrule{2-5}
%     & \textbf{Melodia (Ours)} & \textbf{3.38} & \underline{3.65} & \textbf{3.81} \\
%     \bottomrule
%   \end{tabular}
% \end{table}

\begin{figure}
    \centering
    \includegraphics[width=0.9\linewidth]{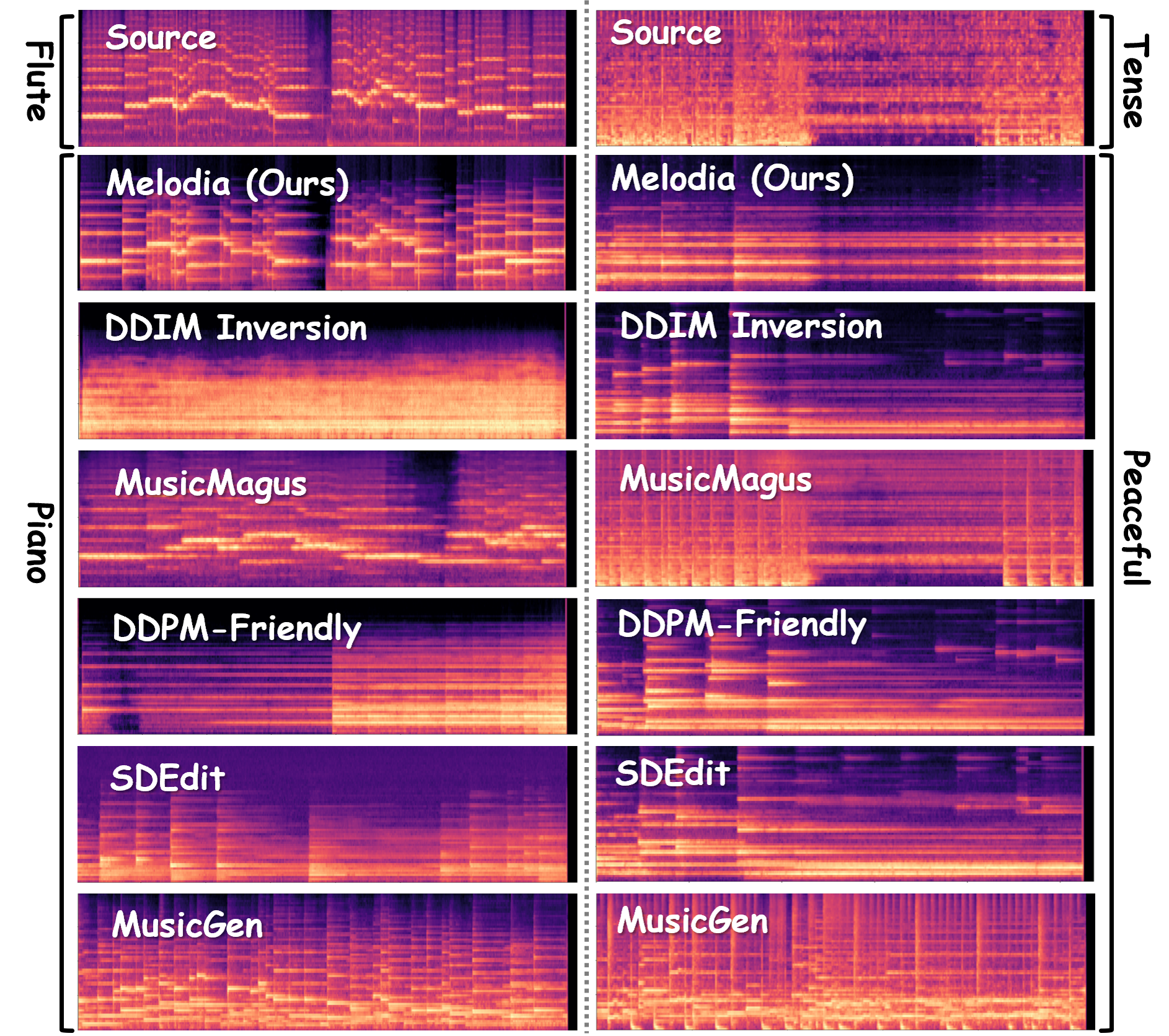}
    \caption{Spectrogram comparison of editing results between Melodia and baseline methods. Melodia successfully preserves the structure and achieves effective attribute transfer. Soundtracks can be found in our demo page.}
    \label{fig:comparison_spect}
\end{figure}

\textbf{Qualitative Evaluation}. Fig.~\ref{fig:comparison_spect} compares editing spectrograms between Melodia and baseline methods. Our method successfully preserves the temporal structure of the original music while effectively transferring the target attributes, which can be verified by more consistent rhythmic patterns maintained from the source music and successful frequency pattern transfer. All methods use the same hyperparameter settings as in the quantitative evaluation. 

\subsection{Subjective Evaluation}
We implemented subjective listening evaluations across three datasets. A total of 100 participants were recruited from the MIR community and music forums, with 77 valid responses retained after screening. All participants had some degree of musical experience. The test is approved by the Institutional Review Board (IRB). We randomly selected 10 test samples from each dataset, and participants rated the edited results from each method. As shown in Table \ref{tab:subjective_comparison}, Melodia substantially outperforms all baseline methods on the MEB metric, indicating that our method successfully achieves the desired balance between textual alignment and structural preservation. Melodia also obtained high scores on REL and CON metrics, confirming its superior performance in semantic alignment and structure preservation.

% \textbf{For more details, please refer to the Appendix.}
% These results confirm that our approach excels in human perceptual evaluation, demonstrating its practical value and real-world applicability. 

\subsection{Additional Analysis}
\textbf{Effects of layer    selection.} The selection of SA replacement layers is crucial in balancing structure preservation and attribute modification. Experimental results on MelodiaEdit demonstrate that replacing SA maps in layers 8–14 of AudioLDM 2 achieves optimal performance. As shown in Tab.~\ref{tab:duoceng}, this layer selection yields the highest balance scores (ASB and AMB) while maintaining low structural disruption (LPAPS) and high textual alignment (CLAP). Visual evidence in Fig. 2 also confirms that SA manipulation in layers 8-14 successfully preserves rhythmic patterns and melodic contours during attribute transfer. 

\textbf{Results on another pre-trained diffusion model.} To evaluate the generalization capability of Melodia across different diffusion models, we conducted experiments with the \textit{Stable Audio Open}~\cite{evans2025stableaudioopen} for timbre transfer on MelodiaEdit. As shown in Tab.~\ref{tab:sd_timbre}, Melodia achieves improvements in both textual adherence (CLAP) and structural integrity (LPAPS and Chroma) compared to the baseline. These results confirm Melodia's generalizability across different model architectures. Additionally, Melodia operates effectively across different sampling rates: 16kHz on AudioLDM2~\cite{audioldm2} and 44.1kHz on Stable Audio Open~\cite{evans2025stableaudioopen}.

\begin{table}[t]
\centering
\footnotesize
\setlength{\tabcolsep}{0.1em}
\caption{Experimental results on Stable Audio Open.}
\label{tab:sd_timbre}
\begin{tabular}{l|cccc}
\hline
\textbf{Method}               & \textbf{CLAP↑} & \textbf{LPAPS↓} & \textbf{Chroma↑} & \textbf{FAD↓} \\ 
\hline
Stable Audio Open             & 0.43           & 6.47            & 0.12             & 1.42          \\
Stable Audio Open + Melodia   & \textbf{0.44}  & \textbf{6.19}    & \textbf{0.21}     & \textbf{1.19} \\
\hline
\end{tabular}
\end{table}

\section{CONCLUSION}
We proposed Melodia, a training-free music editing method achieving optimal balance between textual adherence and structural integrity. Our approach stems from a key insight through attention probing analysis: cross-attention maps encode musical attributes, while self-attention maps preserve temporal structure. By selectively manipulating self-attention maps during denoising, we enable effective editing without requiring textual descriptions of the original music.

Our experimental results demonstrate that Melodia outperforms existing approaches across various datasets. Through our attention repository and Structure Retention technique, we achieved excellent performance in both objective and subjective evaluations. Melodia excels where previous approaches fail, simultaneously maintaining original melody while transforming to target attributes, without additional training or source music descriptions.

\section{Acknowledgments}
This work was supported in part by the National Natural Science Foundation of China under Grant 62202174, in part by the GJYC program of Guangzhou under Grant 2024D01J0081, and in part by the ZJ program of Guangdong under Grant 2023QN10X455, and in part by the Fundamental Research Funds for the Central Universities under Grant 2025ZYGXZR053.

\bibliography{main}

\end{document}